\documentclass[twocolumn,aps,prl,superscriptaddress]{revtex4-2}

\usepackage{amsmath}
\usepackage{graphicx}
\usepackage{MnSymbol}
\usepackage{parskip}
\usepackage{makecell}
\usepackage{booktabs}
\usepackage{multirow}
\usepackage{xcolor}

\makeatletter

\def\section{%
  \@startsection
    {section}%
    {1}%
    {\z@}%
    {0.8cm \@plus 1ex \@minus .2ex}%
    {0.5cm}%
    {\normalfont\small\bfseries\raggedright}%
}

\def\subsection{%
  \@startsection
    {subsection}%
    {2}%
    {\z@}%
    {0.8cm \@plus 1ex \@minus .2ex}%
    {0.5cm}%
    {\normalfont\small\bfseries\raggedright}%
}

\def\subsubsection{%
  \@startsection
    {subsubsection}%
    {3}%
    {\z@}%
    {0.8cm \@plus 1ex \@minus .2ex}%
    {0.5cm}%
    {\normalfont\small\itshape\raggedright}%
}

\makeatother

\begin{document}

\title{Transient fluid removal at soft interfaces: Contact-time-controlled squeeze-out in a cylinder-on-flat contact}

\author{R. Xu}
\affiliation{Peter Gr\"unberg Institute (PGI-1), Forschungszentrum J\"ulich, 52425, J\"ulich, Germany}
\affiliation{State Key Laboratory of Solid Lubrication, Lanzhou Institute of Chemical Physics, Chinese Academy of Sciences, 730000 Lanzhou, China}
\affiliation{MultiscaleConsulting, Wolfshovener str. 2, 52428 J\"ulich, Germany}

\author{B.N.J. Persson}
\affiliation{Peter Gr\"unberg Institute (PGI-1), Forschungszentrum J\"ulich, 52425, J\"ulich, Germany}
\affiliation{State Key Laboratory of Solid Lubrication, Lanzhou Institute of Chemical Physics, Chinese Academy of Sciences, 730000 Lanzhou, China}
\affiliation{MultiscaleConsulting, Wolfshovener str. 2, 52428 J\"ulich, Germany}

\begin{abstract}
We study transient fluid removal in cylinder-on-flat contacts between stiff PMMA
cylinders and a soft PDMS substrate. The cylinder geometry eliminates the edge-scraping
mechanism that can occur for deformable rubber blocks, allowing the influence
of sliding on fluid squeeze-out to be examined more directly. Experiments were performed
mainly in glycerol at the low sliding speed $v=3 \ {\rm \mu m/s}$ using cylinders with different
surface roughness. After the initial elastic-loading stage, we find that the friction during
sliding depends primarily on the total time elapsed since application of the normal load rather
than on the preceding sliding distance. The subsequent friction evolution follows approximately
the same dependence on total contact time. Stationary squeeze-out calculations predict the
evolution of the mean surface separation and real contact area, in reasonable agreement with
that inferred from the measured friction. These results show that essentially the same
squeeze-out process governs fluid removal during stationary contact and low-speed sliding, and
that sliding-induced elastohydrodynamic effects have only a minor influence under these
conditions.
\end{abstract}

\maketitle

\setcounter{page}{1}
\pagenumbering{arabic}

%\pagestyle{empty}

%%%%%%%%%%%%%% main text %%%%%%%%%%%%%%%%
%\begin{multicols}{2}

%%%%%%%%%%%%%% main text %%%%%%%%%%%%%%%%

\section{1 Introduction}

\begin{figure}[tbp]
\includegraphics[width=0.47\textwidth,angle=0]{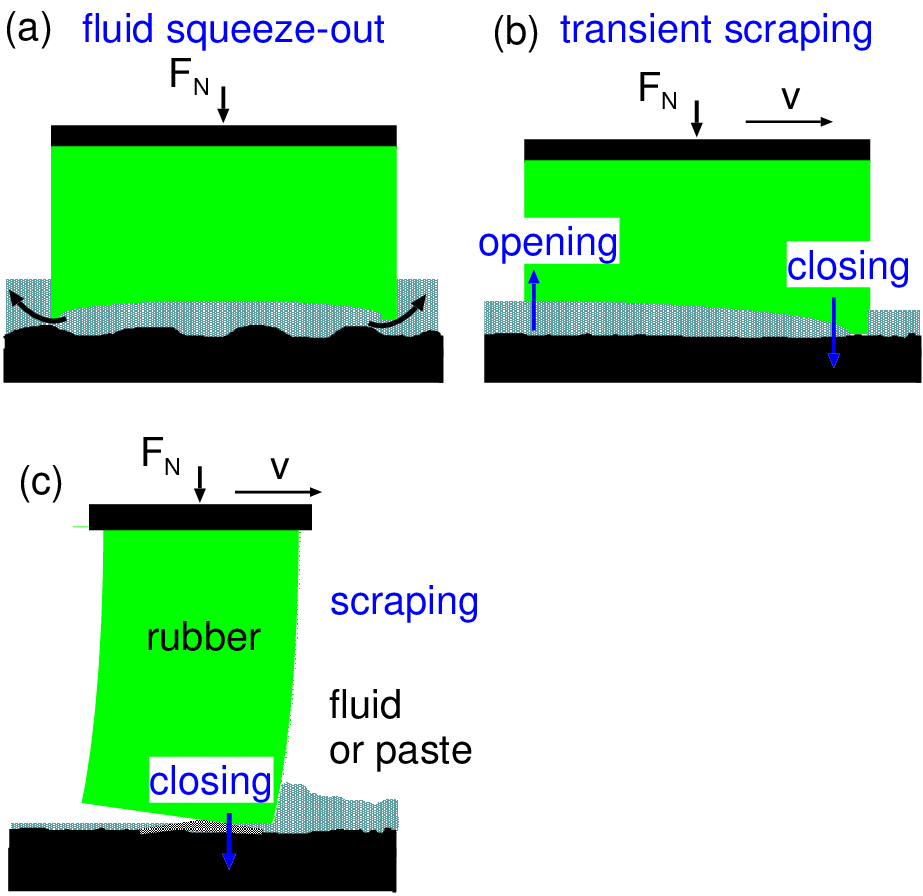}
\caption{\label{Scraping.eps}
Three fluid-removal mechanisms in a block-on-flat contact:
(a) stationary fluid squeeze-out, (b) sliding-induced transient
elastohydrodynamic scraping, and (c) direct leading-edge scraping caused by
bending of the rubber block. The mechanism in (c) is absent in the
cylinder-on-flat geometry considered in the present study. Adapted from
Ref.~\cite{Squeeze1}.}
\label{Scraping.eps}
\end{figure}

The frictional response of lubricated soft contacts during the initial stage of sliding depends
strongly on the evolution of the interfacial fluid film, and many practical contacts remain in this
transient regime rather than reaching steady sliding \cite{Squeeze1}. For example, rubber tread blocks remain
within the tire-road footprint for only a short time and undergo only a limited sliding distance.
The interfacial fluid must therefore be removed rapidly for sufficient friction to develop.
Nevertheless, most studies focus on steady-state friction and the Stribeck curve, which describes
the friction coefficient as a function of sliding speed
\cite{Japan1,Japan2,Japan3,Japan4,Japan5,Japan6}. The mechanisms governing fluid removal
during the transition from stationary contact to sliding are therefore of particular interest.

In a previous study, we investigated the fluid-removal mechanisms in block-on-flat contacts
involving rectangular rubber blocks sliding against tile and glass surfaces in the presence of
water, glycerol, mud, or silicone grease \cite{Squeeze1} (see also Ref. \cite{JCPsqueeze}). 
Fig.~\ref{Scraping.eps} illustrates
three possible fluid-removal mechanisms. In Fig.~\ref{Scraping.eps}(a), fluid is removed by
squeeze-out during stationary contact. In Fig.~\ref{Scraping.eps}(b), sliding produces an
asymmetric fluid pressure distribution, with a negative pressure near the leading edge and a
positive pressure near the trailing edge. The resulting elastohydrodynamic deformation promotes
transient fluid removal. In Fig.~\ref{Scraping.eps}(c), bending of the rubber block brings its
leading edge into contact with the substrate, allowing the fluid to be removed by direct scraping.

The block-on-flat experiments showed that the relative importance of these mechanisms depends
strongly on the fluid properties. For water, stationary squeeze-out was nearly complete before
the onset of sliding. For glycerol, both stationary squeeze-out and sliding-induced fluid removal
were important. For the more viscous mud and grease, the steady-sliding state was reached after
a sliding distance of the order of the block width in the sliding direction, with only a weak
dependence on the preceding waiting time. Thus, sliding-induced scraping was the dominant
fluid-removal mechanism for the highly viscous substances.

In the block-on-flat geometry, the transient elastohydrodynamic process shown in
Fig.~\ref{Scraping.eps}(b) and the edge scraping process shown in
Fig.~\ref{Scraping.eps}(c) may occur simultaneously. Their individual contributions therefore
cannot be clearly separated. A cylinder-on-flat contact does not contain a sharp leading edge
formed by bending of a rubber block, and the mechanism illustrated in
Fig.~\ref{Scraping.eps}(c) is therefore absent. This geometry provides a more direct means of
determining whether sliding itself significantly modifies the rate of fluid squeeze-out through
the process shown in Fig.~\ref{Scraping.eps}(b).
However, the scraping process in Fig.~\ref{Scraping.eps}(b) is an elastohydrodynamic effect and
is important only at sufficiently high sliding speeds. For the cylinder-flat configuration, at high
sliding speeds hydrodynamic lift-off occurs, with the surface separation increasing due to the
fluid pressure. Hence, for the cylinder-flat configuration, the analogue of the transient scraping
process shown in Fig.~\ref{Scraping.eps}(b) is not well defined in most cases.

In the present work, we study polymethyl methacrylate (PMMA) cylinders with different surface roughnesses sliding against
a soft polydimethylsiloxane (PDMS) rubber plate lubricated with glycerol or mud. We investigate the effects of stationary
waiting time, sliding distance, total contact time, sliding speed, and cylinder surface roughness
on the transient friction response. At low sliding speed in glycerol, the interfacial state is found
to depend primarily on the total contact time, indicating that sliding-induced fluid removal has only a minor influence on fluid removal in the cylinder-on-flat contact. At a higher sliding speed in mud, 
hydrodynamic re-entrainment becomes important. We also discuss the effects of surface roughness, 
adhesion, and interfacial instabilities on the transient friction behavior.

\vskip 0.3cm
\section{2 Experimental setup}

Measurements were performed for PMMA cylinders sliding against a silicone rubber plate. The
silicone rubber was prepared using Sylgard 184, a two-component kit purchased from Dow Corning
(Midland, MI), consisting of a base containing vinyl-terminated polydimethylsiloxane and a curing
agent (cross-linker) containing a methylhydrosiloxane-dimethylsiloxane copolymer and a suitable
catalyst. The two components were mixed at a cross-linker-to-base mass ratio of $1:10$. The
mixture was degassed and subsequently cured in an aluminum container at room temperature for
$3$ days. The PDMS rubber had a low-frequency Young's modulus of
$E=1.5 \ {\rm MPa}$ and a Poisson ratio of $\nu=0.5$.

The PMMA cylinders had radii of either $R=7 \ {\rm cm}$ or $R=3 \ {\rm cm}$ and an axial
length of $L=10 \ {\rm cm}$. The cylinders were subjected to the same surface treatments as
in Ref.~\cite{XuPersson}, resulting in different roughness levels. Surfaces sandblasted for
$1 \ {\rm min}$ at an air pressure of $1 \ {\rm MPa}$ were denoted sandblasted-rough (Sb-R).
Surfaces sandblasted for $0.5 \ {\rm min}$ at an air pressure of $0.5 \ {\rm MPa}$ were
denoted sandblasted-smooth (Sb-S). Untreated surfaces were denoted Smooth.

The surface roughness of all contacting surfaces was characterized using a Mitutoyo Surftest
SJ-410 portable surface roughness profilometer equipped with a diamond stylus having a tip radius
of curvature of $2 \ {\rm \mu m}$ and a stylus-substrate loading force of
$0.75 \ {\rm mN}$. The sampling interval (pixel size) was $0.5 \ {\rm \mu m}$, the scan
length was $25 \ {\rm mm}$, and the stylus speed was $50 \ {\rm \mu m/s}$. Assuming
isotropic surface roughness, the 2D power spectra presented below were calculated from the 1D
power spectra obtained by averaging three surface-profile measurements \cite{XuP}.

The interfaces were lubricated with glycerol with a purity of 99.5\%,
or mud. Fresh glycerol was used for each measurement to avoid a reduction in
viscosity caused by the absorption of water from the atmosphere. The mud was prepared by
gradually adding bentonite powder to water at an approximate water-to-bentonite powder volume
ratio of $4:3$. Bentonite powder strongly absorbs water and forms a non-Newtonian,
shear-thinning suspension. The mixture was allowed to hydrate and was subsequently stirred
thoroughly until a macroscopically homogeneous suspension without visible dry-powder
agglomerates was obtained.

The friction experiments were performed using the same linear friction tester as in
Ref.~\cite{Squeeze1}. The normal force could be varied from $F_{\rm N}=31 \ {\rm N}$ to
approximately $1000 \ {\rm N}$, and the sliding velocity could be varied from
$1 \ {\rm \mu m/s}$ to $1 \ {\rm cm/s}$. During the experiments, the PDMS rubber plate,
together with its aluminum container, was attached to the machine table, which was translated
by a servo motor through a gearbox. For the waiting-time experiments, the normal load was first applied and the contact was kept
stationary for a prescribed waiting time before motion of the machine table was
initiated. The total contact time $t$ is defined as the time elapsed since application of the
normal load and therefore includes both the stationary waiting period and the subsequent period
after the onset of table motion.

%%%%%%%%%%%%%%%%%%%%%%%%%%%%%%%%%%%%%%%%%%%%%%%%%%%

\begin{table}[hbt]
   \caption{Cylinder radius $R$, normal force per unit length $F_{\rm N}/L$, Hertzian contact
   width $w$, indentation depth $\delta$, and mean Hertzian contact pressure $p_0$.}
   \label{loading}
   \renewcommand{\arraystretch}{1.5}
   \begin{center}
      \begin{tabular}{@{}|l||c|c|c|c|@{}}
         \hline
            $R \ ({\rm cm})$ & $F_{\rm N}/L \ {\rm (N/m)}$ & $w \ ({\rm cm})$
            & $\delta \ ({\rm mm})$ & $p_0 \ ({\rm MPa})$ \\
         \hline
         \hline
            7.0 & 310  & 0.74 & 0.20 & 0.042 \\
         \hline
            7.0 & 910  & 1.27 & 0.58 & 0.071 \\
         \hline
            3.0 & 910  & 0.83 & 0.58 & 0.109 \\
         \hline
            3.0 & 3600 & 1.66 & 2.29 & 0.217 \\
         \hline
      \end{tabular}
   \end{center}
\end{table}
%
%\end{widetext}

\vskip 0.3cm
\section{3 Theory}

We analyze the experimental results presented below using the fluid squeeze-out theory for the
block-on-flat geometry described in detail in Ref.~\cite{Squeeze1}. Here, we summarize only the
elements needed for the present cylinder-on-flat geometry. The theory considers fluid squeeze-out
between a randomly rough elastic substrate and a rigid rectangular block of width $w=2a$ in the
$x$-direction and infinitely long in the $y$-direction. In Ref.~\cite{Persson1}, it was shown that
this treatment also provides a good approximation of squeeze-out in a cylinder-on-flat contact
when the block width $w$ is identified with the Hertzian contact width. The block approximation
neglects the macroscopic variation of the interfacial gap associated with the curved cylinder
geometry, but, as shown in Ref.~\cite{Persson1}, it provides nearly the same squeeze-out behavior
as the full cylinder-contact calculation for the conditions considered here.

Using the Hertz contact theory for the dry cylinder-on-flat contact, the contact width $w$, indentation
depth $\delta$, and mean contact pressure $p_0$ are given by
$$
w={4\over \surd\pi}
\left({RF_{\rm N}\over E^*L}\right)^{1/2},
\ \ \ 
\delta={w^2\over4R},
\ \ \ 
p_0={\surd\pi\over4}
\left({E^*F_{\rm N}\over RL}\right)^{1/2},
$$
where $E^*=E/(1-\nu^2)$ is the effective elastic modulus of the PDMS rubber, with the PMMA
cylinder treated as rigid. Using the measured low-frequency modulus of the PDMS rubber,
$E=1.5 \ {\rm MPa}$, Table~\ref{loading} summarizes the Hertzian contact parameters for the
systems studied below.

We approximate the cylinder by an infinitely long rigid rectangular block with a width
$w=2a$ in the fluid flow direction, in contact with the elastic PDMS substrate. We introduce an
$xy$-coordinate system in the nominal contact plane, with the origin at the center of the nominal
contact region. The $x$-axis is oriented along the fluid-flow direction, which corresponds to the
sliding direction in the experiments considered below, while the $y$-axis is oriented perpendicular
to the sliding direction. The boundaries of the nominal contact region are located at
$x=\pm a$.

Let $\bar u(x,t)$ and $\bar p(x,t)$ denote the ensemble-averaged interfacial separation and fluid
pressure, respectively. If the mean separation $\bar u$ is assumed to depend only on time, the
effective Reynolds thin-film equation becomes
$${\partial \bar u\over\partial t}
-{\bar u^3\phi_{\rm p}(\bar u)\over 12\eta}
{\partial^2\bar p\over\partial x^2}=0\eqno(1)$$
where $\eta$ is the fluid viscosity and $\phi_{\rm p}(\bar u)$ is the pressure flow factor, which
accounts for the influence of surface roughness on the effective fluid-flow conductivity.

Assuming that the fluid pressure vanishes at the boundaries $x=\pm a$, the pressure distribution
is
$$
\bar p(x,t)=p_{\rm fluid}(t){3\over2}
\left[
1-\left({x\over a}\right)^2
\right]\eqno(2)
$$
where $p_{\rm fluid}(t)$ is the spatially averaged fluid pressure. Substituting (2) into (1) gives
$$
{d\bar u\over dt}
=-{\bar u^3\phi_{\rm p}(\bar u)\over 4\eta a^2}
p_{\rm fluid}(t).\eqno(3)
$$

If $p_0$ is the externally applied nominal pressure acting on the approximated PMMA block, force balance gives
$$
p_{\rm fluid}(t)=p_0-p_{\rm cont}(t),\eqno(4)
$$
where $p_{\rm cont}$ is the ensemble-averaged asperity contact pressure. The relation between $\bar u$ and $p_{\rm cont}$ can be obtained from
Persson contact mechanics theory. At large interfacial separations, this relation takes the form
$$
p_{\rm cont}=\beta E^*e^{-\bar u/u_0},\eqno(5)
$$
where $\beta$ and $u_0$ are determined by the surface roughness power spectrum. Typically,
$u_0\approx h_{\rm rms}$, where $h_{\rm rms}$ is the root-mean-square surface roughness.

Equations (3)-(5), together with the calculated pressure flow factor
$\phi_{\rm p}(\bar u)$, determine the time evolution of the mean interfacial separation
$\bar u(t)$, the fluid pressure $p_{\rm fluid}(t)$, and the asperity contact pressure
$p_{\rm cont}(t)$.

The treatment presented above neglects the macroscopic deformation of the elastic body. As shown in Ref. \cite{Persson1} 
at sufficiently long stationary-contact
times, this bending becomes small and has a negligible influence on the final stages of the
fluid-removal process. 

%%%%%%%%%%%%%%%%%%%%%%%%%%%%%%%%%%%%%%%%%%%%%%%%%%%
\vskip 0.1cm
\section{4 Cylinder-on-flat contact}

We investigate how fluid removal from the interface affects the friction force. This involves
non-stationary sliding, and we denote the ratio between the friction force $F_{\rm f}$ and the
normal force $F_{\rm N}$ by $\mu$, even though this symbol is usually used to denote either the
steady-state kinetic friction coefficient $\mu_{\rm k}$ or the break-loose, or static, friction
coefficient $\mu_{\rm s}$. Thus, $\mu(t)=F_x(t)/F_{\rm N}$ generally depends on time.

As discussed above, in contrast to the
block-on-flat geometry, the cylinder-on-flat contact does not permit the edge scraping
mechanism illustrated in Fig.~\ref{Scraping.eps}(c). It therefore provides a means of examining
more directly whether sliding modifies fluid removal through the transient elastohydrodynamic
mechanism illustrated in Fig.~\ref{Scraping.eps}(b).

\begin{figure}[tbp]
\includegraphics[width=0.47\textwidth,angle=0]{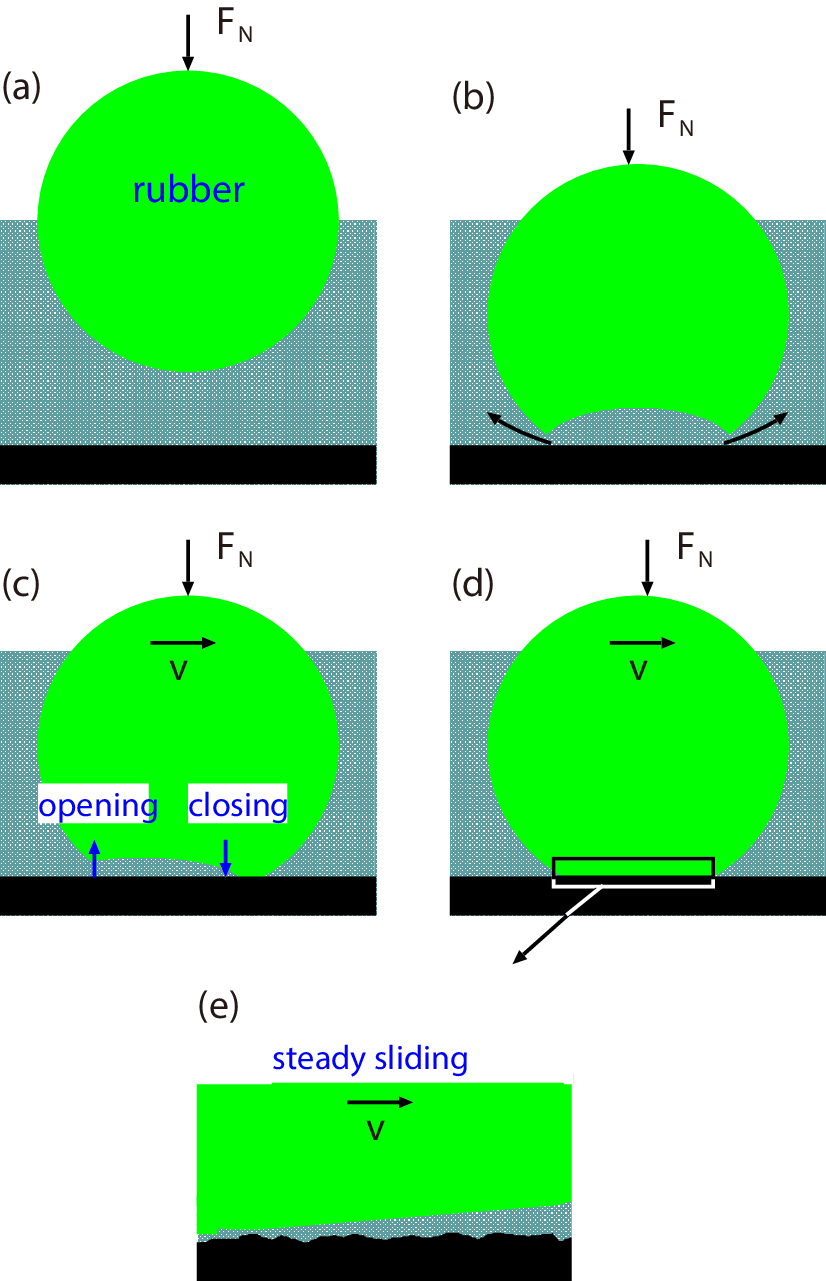}
\caption{\label{BallSlidingPicture.eps}
Schematic illustration of sliding-induced fluid removal through elastohydrodynamic deformation.
(a) Initial configuration before application of the normal load. (b) During stationary normal
contact, the fluid pressure deforms the elastic surface. (c) and (d) During the subsequent sliding
motion, negative and positive fluid-pressure regions develop near the leading and trailing edges,
respectively, resulting in opening and closing of the interfacial gap. Depending on the applied
tangential force, the final state may be either a no-slip state or steady sliding accompanied by
elastohydrodynamic deformation, as illustrated in (e). Although an elastic cylinder on a rigid flat
surface is shown, an analogous mechanism can occur for a rigid cylinder sliding on an elastic
half-space.
}
\end{figure}

Fig.~\ref{BallSlidingPicture.eps} schematically illustrates this process for an elastic cylinder
in contact with a rigid flat surface. During stationary normal loading, the fluid pressure deforms
the elastic surface. When sliding begins, an asymmetric fluid-pressure distribution may develop,
with a negative pressure near the leading edge and a positive pressure near the trailing edge.
The resulting opening and closing of the interfacial gap can promote fluid removal during sliding.
Although the schematic shows an elastic cylinder and a rigid flat surface, an analogous evolution
of the interfacial gap can occur for the rigid PMMA cylinder sliding on the elastic PDMS substrate
used in the present experiments, with the deformation occurring primarily in the PDMS.

In Figs.~\ref{BLOCK.1distance.2mu.Sb-S.glycerol.two.waitinngTimes.eps}-
\ref{BLOCK.1Logt.2mu.Smooth.replot.eps}, the friction coefficient is shown as a function of
sliding distance or time using the original time resolution of $10$ data points per second.
In all other figures, the friction coefficients were averaged over very narrow sliding-distance
intervals ($0.1 \ {\rm mm}$ or less) in order to reduce the noise in the measured data.

Below, we present friction results for the dry contact and for contacts lubricated with glycerol.
In Appendices A and B, we present additional results for the Sb-S cylinder in glycerol and in mud, respectively.

\vskip 0.1cm
\subsection{4.1 Dry-contact response and interfacial instabilities}

\begin{figure}
\includegraphics[width=1.0\columnwidth]{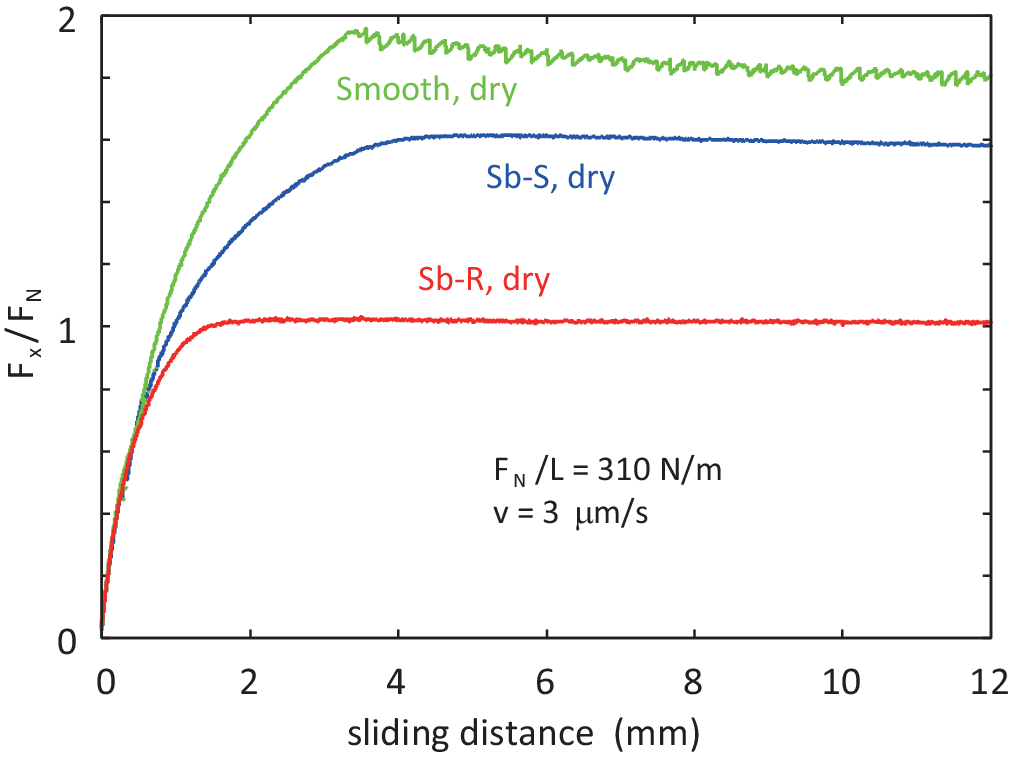}
\caption{\label{1SlidingDistance.2Mu.smooth.dry.lubricated.eps}
The friction coefficient as a function of sliding distance. The green, blue, and red
curves correspond to the Smooth, Sb-S, and Sb-R PMMA surfaces, respectively. The sliding speed is
$v=3 \ {\rm \mu m/s}$, and the normal force per unit length is
$F_{\rm N}/L=310 \ {\rm N/m}$. The measured data were averaged over sliding-distance intervals
of $0.1 \ {\rm mm}$ to reduce random statistical noise.
}
\end{figure}

\begin{figure}
\includegraphics[width=1.0\columnwidth]{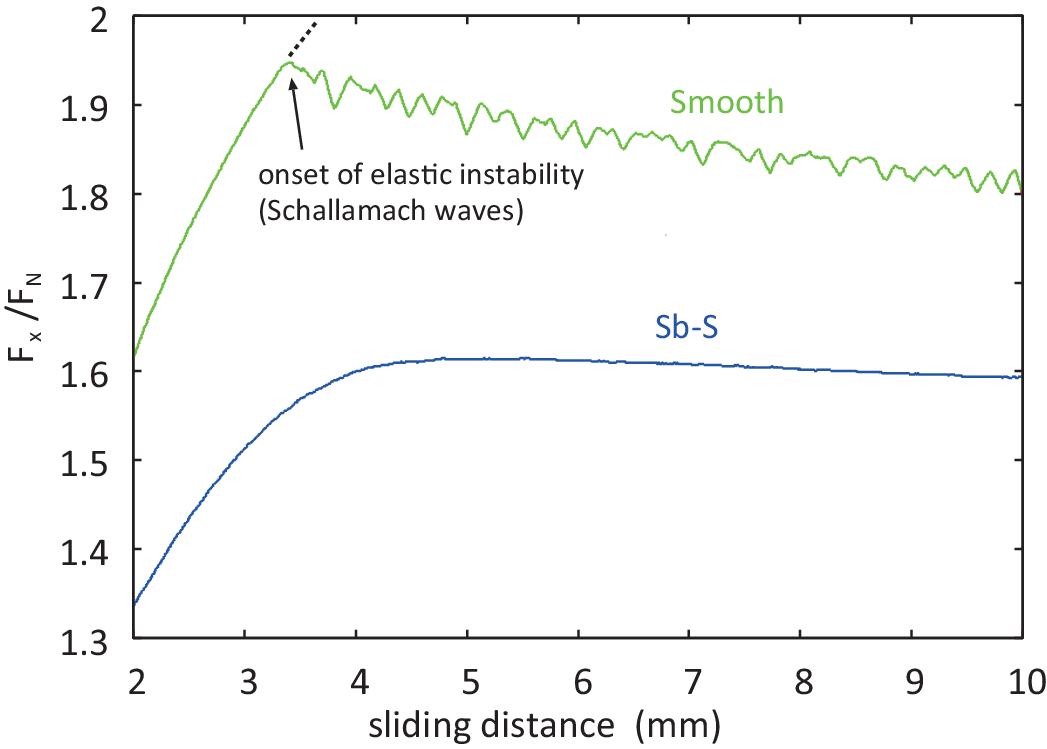}
\caption{\label{1x.2mu.Smooth.Sb-S.dry.MAGNIFIED.eps}
Magnified view of the initial sliding-distance range in
Fig.~\ref{1SlidingDistance.2Mu.smooth.dry.lubricated.eps}. The green and blue curves correspond
to the Smooth and Sb-S PMMA surfaces, respectively.
}
\end{figure}

Fig.~\ref{1SlidingDistance.2Mu.smooth.dry.lubricated.eps} shows the friction coefficient as a function
of sliding distance under dry-contact conditions for the Smooth, Sb-S, and Sb-R.
At very small imposed displacements, there is no global slip at the interface. The initial slope
of each curve is therefore determined by the tangential stiffness of the contact, which depends
on the contact-stress distribution and the elastic modulus of the rubber; see Appendix C.
At somewhat larger displacements, partial slip, or preslip, may occur before the onset of global
sliding. In this regime, some regions of the interface remain pinned while slip occurs elsewhere.

Fig.~\ref{1x.2mu.Smooth.Sb-S.dry.MAGNIFIED.eps} shows a magnified view of the initial sliding distance
range in Fig.~\ref{1SlidingDistance.2Mu.smooth.dry.lubricated.eps}. The green and blue curves correspond
to the Smooth and Sb-S surfaces, respectively.

For the Smooth, the friction initially evolves smoothly with increasing sliding distance
until elastic instabilities (Schallamach waves) appear. By contrast, the friction curves for the Sb-S and Sb-R
remain smooth over the full sliding-distance range. In Ref.~\cite{XuPersson}, we showed that the
Smooth has a nonzero macroscopic effective interfacial binding energy, 
both in the dry state and in glycerol. This is not the case for
the Sb-S and Sb-R, for which surface roughness suppresses (or kill) the macroscopic adhesion.

Therefore, for the Smooth, adhesion can act over length scales comparable to the nominal
contact region and generate macroscopic elastic instabilities. This explains
the stick-slip-like features in the friction curve shown in
Fig.~\ref{1x.2mu.Smooth.Sb-S.dry.MAGNIFIED.eps}. To illustrate this point, it is useful to consider
what would occur in the absence of Schallamach waves. If sliding occurred uniformly over the entire
interface, and if the effective interfacial shear stress were similar to that of the Sb-S,
then the friction coefficient of the Smooth would continue to increase with sliding distance,
as indicated by the dashed curve in
Fig.~\ref{1x.2mu.Smooth.Sb-S.dry.MAGNIFIED.eps}, and could reach values of the order of $\sim 5$.

However, when the shear stress in the nominal contact region becomes comparable to the Young's modulus
of the rubber, elastic instabilities may occur. Detached regions of the rubber surface then
propagate through the contact in the form of Schallamach waves \cite{Schallwaves,chaud}. The resulting
motion is no longer governed by uniform interfacial slip throughout the nominal contact region,
but instead by instability-mediated propagation, which can produce a much lower effective friction
coefficient.

\vskip 0.1cm
\subsection{4.2 Fluid squeeze-out in glycerol}

The Stribeck curves for the cylinders sliding on the glycerol-covered silicone rubber surface
were studied in Ref.~\cite{XuPersson}. Most of the experiments considered here were performed
at the low sliding speed $v=3 \ {\rm \mu m/s}$. At this speed, the steady-state contacts involving
the Sb-S and Sb-R surfaces are in the boundary-lubrication regime of the Stribeck curve. The
Smooth surface exhibits additional adhesion-controlled effects and interfacial instabilities, which
are considered separately below.

Separate from these adhesion-controlled effects, sliding may modify fluid removal through an
elastohydrodynamic pressure redistribution. In particular, a reduction in fluid pressure near the
entrance of the contact may contribute to closing the interfacial gap, as illustrated in
Figs.~\ref{Scraping.eps}(b) and \ref{BallSlidingPicture.eps}(c). However, this hydrodynamic effect is expected to be weak at low
sliding velocities. At $v=3 \ {\rm \mu m/s}$, the sliding-induced pressure redistribution should
therefore have only a minor influence, and fluid removal during sliding is expected to proceed at approximately
the same rate as stationary squeeze-out [Fig.~\ref{Scraping.eps}(a)]. The experiments presented
below investigate fluid squeeze-out under these conditions and test whether the interfacial state
is governed primarily by the total contact time rather than by the sliding distance, as expected
if sliding-induced hydrodynamic effects are negligible.

\begin{figure}
\includegraphics[width=1.0\columnwidth]{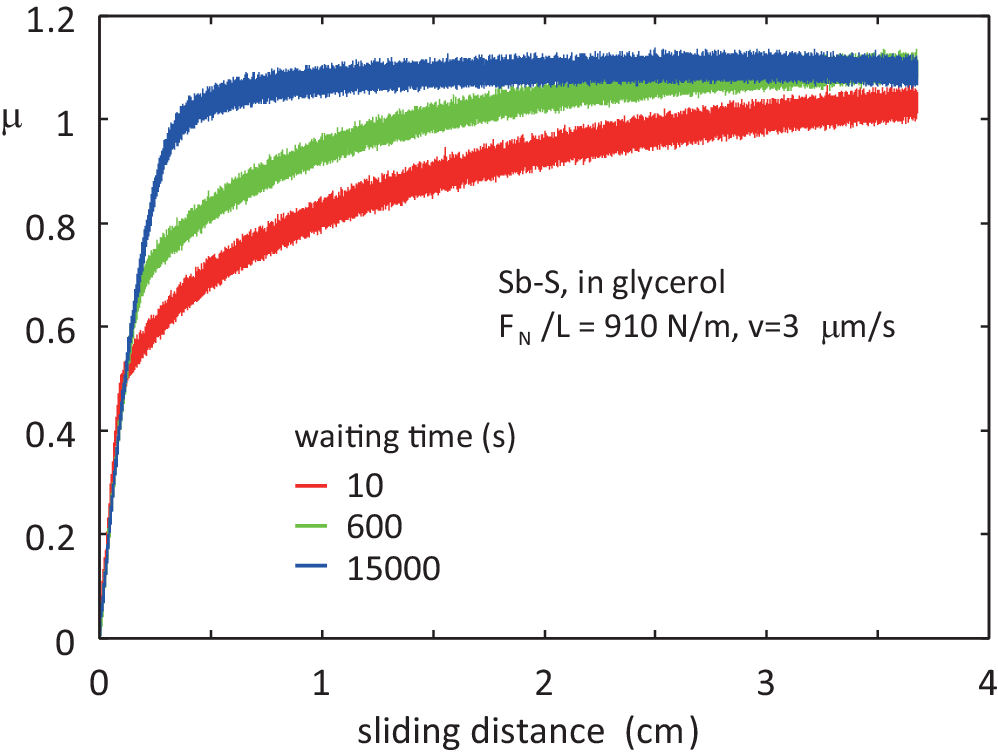}
\caption{\label{BLOCK.1distance.2mu.Sb-S.glycerol.two.waitinngTimes.eps}
The friction coefficient for the Sb-S cylinder sliding against a flat PDMS substrate in glycerol
as a function of sliding distance. The red, green, and blue curves correspond to waiting times
$t_{\rm w}=10 \ {\rm s}$, $10 \ {\rm min}$, and $4.2 \ {\rm h}$, respectively. The sliding
speed is $v=3 \ {\rm \mu m/s}$, and the normal force per unit length is
$F_{\rm N}/L=910 \ {\rm N/m}$.
}
\end{figure}

\begin{figure}
\includegraphics[width=1.0\columnwidth]{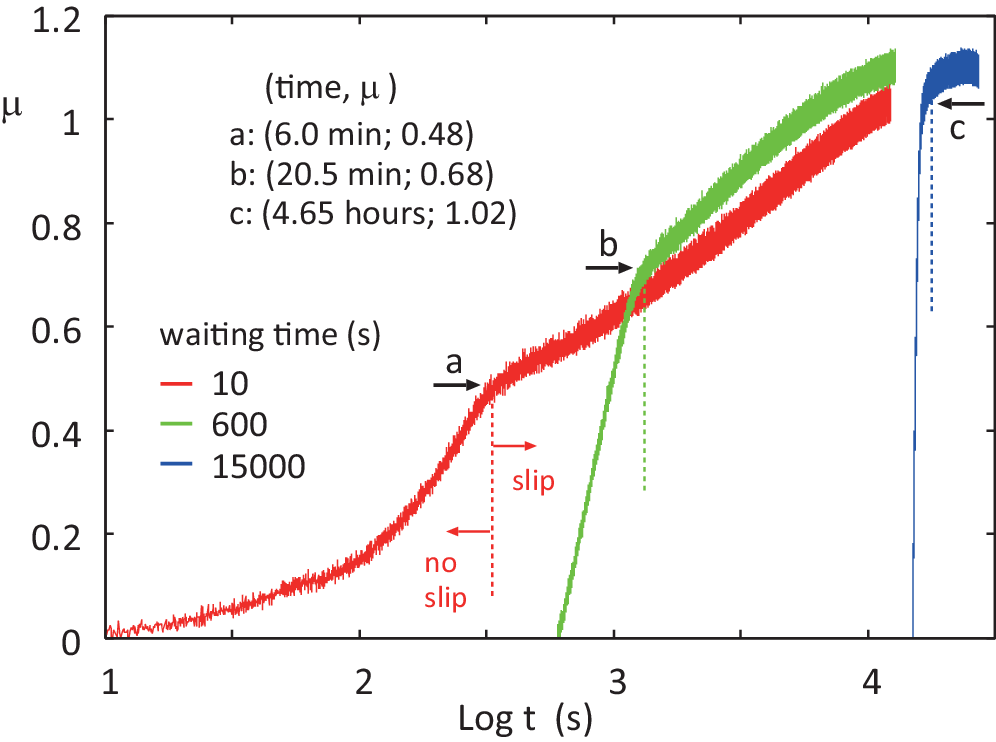}
\caption{\label{1Logt.2mu.NEW3cm.Sb-S.glycerol.2plates.eps}
The same data as in
Fig.~\ref{BLOCK.1distance.2mu.Sb-S.glycerol.two.waitinngTimes.eps}, plotted as functions
of the total contact time elapsed since application of the normal load. The points
${\bf a}$, ${\bf b}$, and ${\bf c}$ indicate the onsets of global sliding. The increase in
friction up to these points is caused primarily by tangential elastic deformation of the contact,
although some local preslip may occur. The regions to the left of the vertical dashed lines
correspond to the no-slip or partial-slip portions of the friction curves.
}
\end{figure}

Fig.~\ref{BLOCK.1distance.2mu.Sb-S.glycerol.two.waitinngTimes.eps} shows the friction coefficient
for the Sb-S cylinder sliding against a flat PDMS substrate in glycerol as a function of sliding
distance at $v=3 \ {\rm \mu m/s}$. The red, green, and blue curves correspond to waiting times
of $10 \ {\rm s}$, $10 \ {\rm min}$, and $4.2 \ {\rm h}$, respectively.

Fig.~\ref{1Logt.2mu.NEW3cm.Sb-S.glycerol.2plates.eps} shows the same data plotted as functions
of the total contact time. As indicated by the dry-contact measurements and the contact-stiffness
analysis in Appendix C, the initial increase in friction after the onset of driving is governed
mainly by tangential elastic deformation of the contact. Some local preslip may occur during this
stage, but global sliding does not begin until the points indicated by the arrows
${\bf a}$, ${\bf b}$, and ${\bf c}$.

At point ${\bf b}$, the friction coefficient at the onset of global sliding after a waiting time
of $10 \ {\rm min}$ is close to the value reached by the red curve at the same total contact time.
Likewise, at point ${\bf c}$, the friction coefficient after a waiting time of $4.2 \ {\rm h}$ is
close to the extrapolated values of the red and green curves at the same total contact time. Once
global sliding begins, the curves continue to follow approximately the same dependence on total
contact time, despite their different waiting times and sliding distances. In particular, the
friction coefficient increases from $\mu\approx0.1$ at $t\approx100 \ {\rm s}$ to
$\mu\approx1$ at $t\sim10^4 \ {\rm s}$, close to the steady-state friction coefficient.

These results indicate that, for the cylinder-on-flat geometry at
$v=3 \ {\rm \mu m/s}$, the interfacial state is governed primarily by the total time elapsed
since application of the normal load rather than by the preceding sliding distance. Thus,
sliding-induced fluid removal [process (b) in Fig. \ref{Scraping.eps} and (c) in Fig. \ref{BallSlidingPicture.eps}] 
has at most a minor influence on the squeeze-out rate under the
present conditions. This is due to the low sliding speed, which makes elastohydrodynamic effects
of the type shown in Fig.~\ref{BallSlidingPicture.eps}(c) unimportant.
This interpretation is examined quantitatively in Sec.~4.3 by comparison
with stationary squeeze-out theory.

\begin{figure}
\includegraphics[width=1.0\columnwidth]{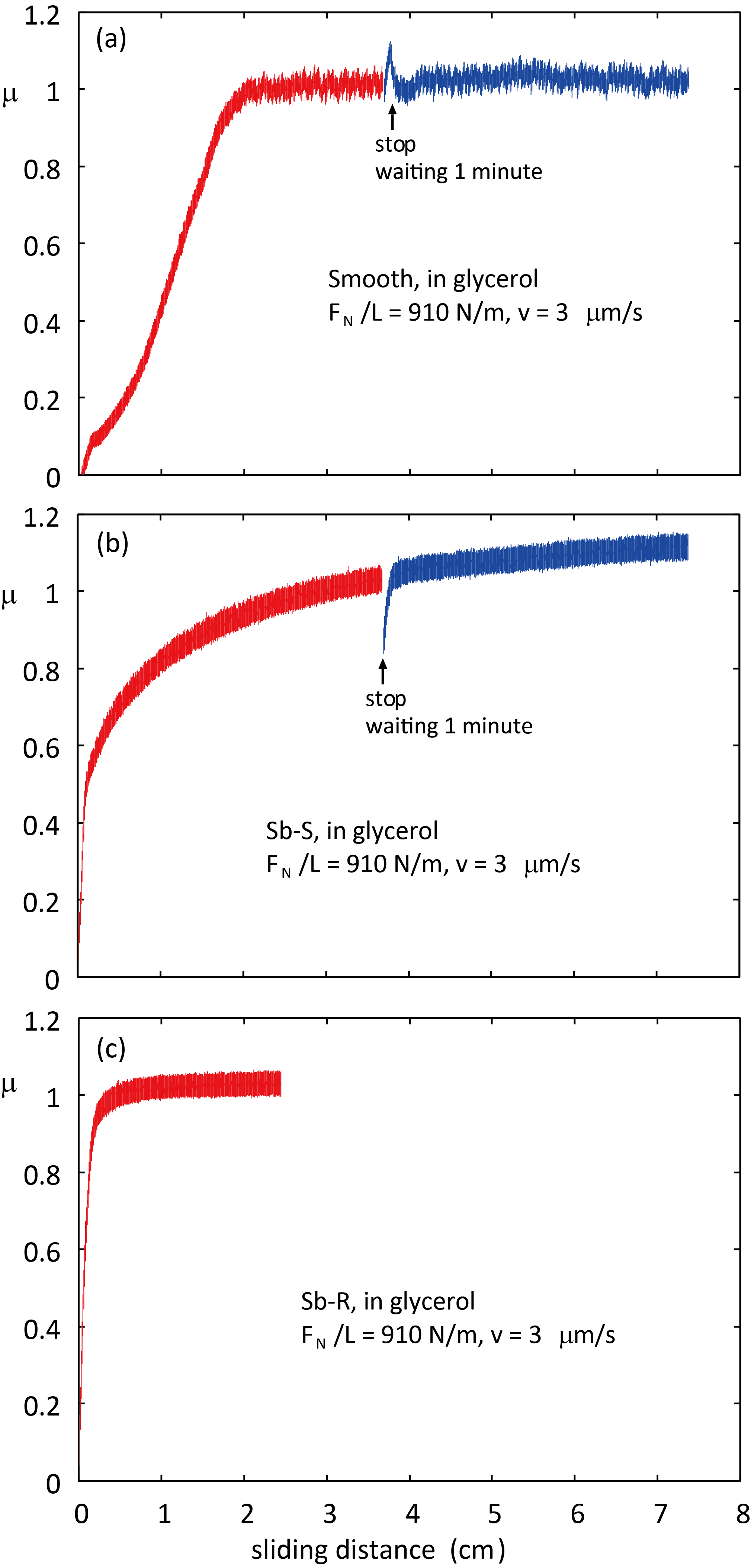}
\caption{\label{SCRAPING.1x.2mu.eps}
The friction coefficient of (a) the Smooth, (b) the Sb-S, and (c) the Sb-R cylinder sliding
against a flat PDMS substrate in glycerol as a function of sliding distance. For cases (a) and (b), the cylinder was first
squeezed against the PDMS surface for $10 \ {\rm s}$ and was then abruptly driven at
$v=3 \ {\rm \mu m/s}$. After a sliding distance of approximately $2 \ {\rm cm}$, the motion was
stopped for $1 \ {\rm min}$ and was subsequently restarted abruptly at
$v=3 \ {\rm \mu m/s}$. The normal force per unit length was
$F_{\rm N}/L=910 \ {\rm N/m}$.
%The calculated Hertzian contact width in the sliding direction is $w=1.27 \ {\rm cm}$.
}
\end{figure}

The stop-restart experiments shown in Fig.~\ref{SCRAPING.1x.2mu.eps} further supported this conclusion. The red curves correspond to the first sliding stage, during which the
cylinders were driven at the constant speed $v=3 \ {\rm \mu m/s}$ over a distance of
approximately $2 \ {\rm cm}$. The motion was then stopped, and the system was held stationary
for approximately $1 \ {\rm min}$. The drive velocity was subsequently restored to
$v=3 \ {\rm \mu m/s}$, corresponding to the blue curves.

Comparison of Fig.~\ref{1SlidingDistance.2Mu.smooth.dry.lubricated.eps} with
Fig.~\ref{SCRAPING.1x.2mu.eps}(a) shows that the increase in friction with sliding distance for
the Smooth cylinder is much slower in the lubricated contact than in the dry contact, consistent
with the gradual removal of glycerol from the nominal contact region. However, the evolution of
the Smooth contact cannot be interpreted solely in terms of fluid squeeze-out. As shown in
Ref.~\cite{XuPersson}, the Smooth surface has a nonzero effective interfacial binding energy and therefore
exhibits macroscopic adhesion, whereas surface roughness suppresses macroscopic adhesion for the
Sb-S and Sb-R surfaces.

The nonzero effective interfacial binding energy of the Smooth surface promotes forced dewetting once
asperity contacts begin to form. This explains why in Figs.~\ref{SCRAPING.1x.2mu.eps}(a)-(c), the Smooth contact reaches its apparent
steady-state friction level over a shorter sliding distance than the Sb-S contact, but still longer
than that for the Sb-R surface. For the Sb-R surface, the higher roughness leads to faster
squeeze-out and earlier formation of asperity contacts because asperity contact begins at a larger
mean interfacial separation.

When sliding is resumed during the second sliding stage, the friction starts from approximately
the value reached at the end of the first sliding stage. This behavior is consistent with the
conclusion that the interfacial state depends primarily on the total contact time and is not reset
by stopping and restarting the sliding motion. For the Sb-S surface, the restarted curve exhibits
approximately the same elastic slope as the initial loading curve. For the Smooth surface, a
transient friction peak occurs after sliding is resumed. We attribute this peak to the formation
of additional interfacial bonds during the $1 \ {\rm min}$ stationary period and the subsequent
rupture of these bonds upon restarting.

\begin{figure}
\includegraphics[width=1.0\columnwidth]{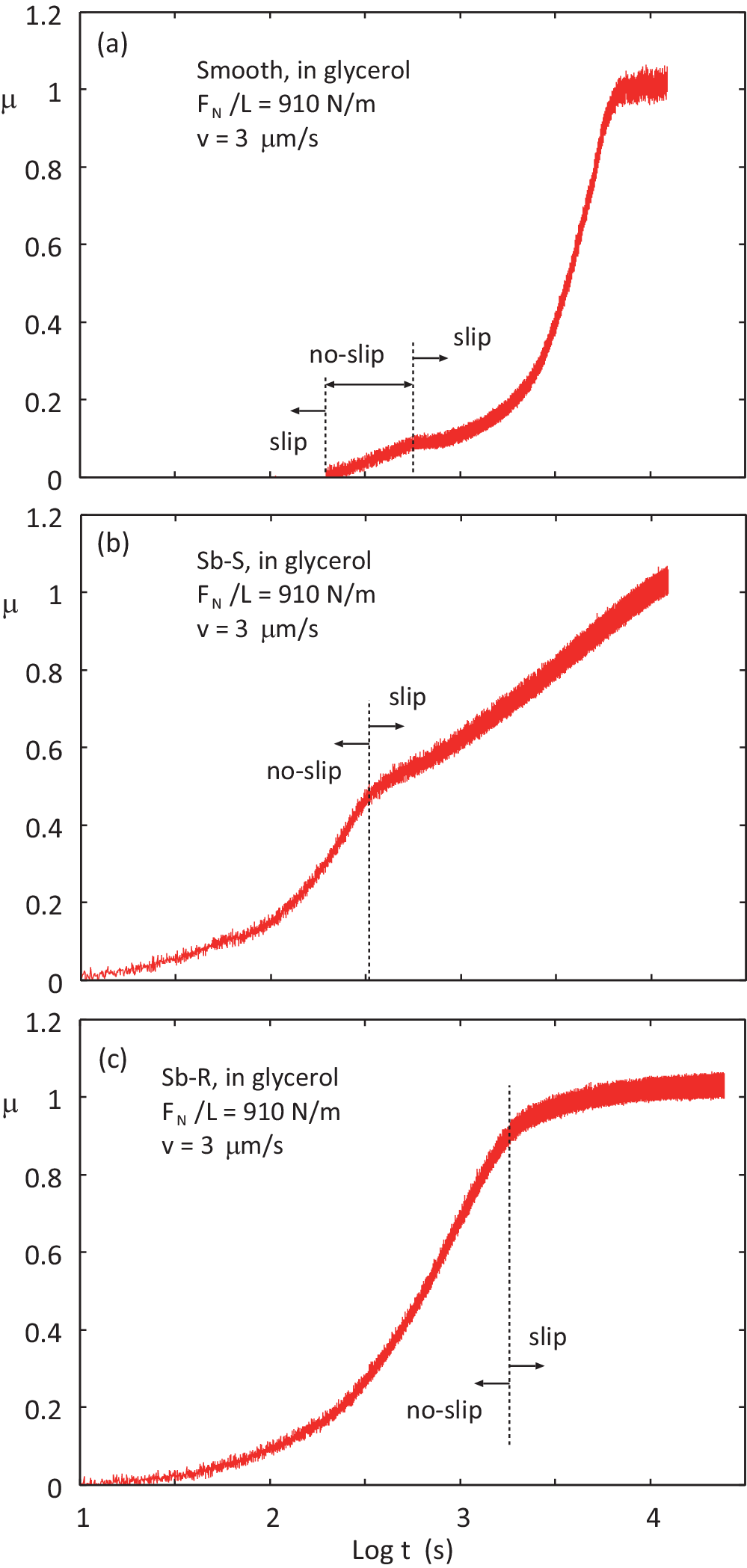}
\caption{\label{BLOCK.1Logt.2mu.Smooth.replot.eps}
The friction coefficient during the first sliding stage as a function of the logarithm of the
total contact time elapsed since application of the normal load. The data correspond to the red
curves in Fig.~\ref{SCRAPING.1x.2mu.eps}. Panels (a), (b), and (c) show the results for the
Smooth, Sb-S, and Sb-R PMMA cylinders, respectively.
}
\end{figure}

To examine how surface roughness and adhesion influence the transient friction response,
Fig.~\ref{BLOCK.1Logt.2mu.Smooth.replot.eps} shows the first sliding stage from
Fig.~\ref{SCRAPING.1x.2mu.eps} as a function of the total contact time.

Fig.~\ref{BLOCK.1Logt.2mu.Smooth.replot.eps}(a) shows the results for the Smooth surface.
For $t\lesssim150 \ {\rm s}$, the fluid film is sufficiently thick that the real asperity
contact area is negligible. The resulting friction is comparable to the instrumental resolution, and is not shown in the figure.

For $150 \ {\rm s}\lesssim t\lesssim500 \ {\rm s}$, sufficient fluid has been removed for
asperity contact to begin. The interfacial tangential stiffness then becomes large enough for the
contact to enter a transient no-slip or partial-slip state. For $t\gtrsim500 \ {\rm s}$, global
sliding occurs and the friction continues to increase as the interfacial state evolves. This
evolution involves further fluid removal together with the adhesion-controlled dewetting discussed
above. The apparent steady-state friction level is reached for $t\gtrsim10^4 \ {\rm s}$. As
shown in Sec.~4.3, the calculated real contact area is negligible for
$t<150 \ {\rm s}$, consistent with the experimental behavior in
Fig.~\ref{BLOCK.1Logt.2mu.Smooth.replot.eps}(a).

Fig.~\ref{BLOCK.1Logt.2mu.Smooth.replot.eps}(b) shows the corresponding results for the Sb-S
surface. The rapid initial increase in friction up to ${\rm Log} \, t\approx 2.5$, corresponding to
$t\approx300 \ {\rm s}$, is mainly caused by tangential elastic deformation of the contact before
the onset of global slip. Some local preslip may nevertheless occur and influence the slope of the
force-displacement curve.

The pronounced elastic-loading stage indicates that a finite asperity contact area is already
present at the onset of motion of the machine table, and that the surfaces at the sliding interface
remain pinned until the tangential force reaches approximately $40\%$ of the steady-state sliding
friction force. Thus, the greater roughness of
the Sb-S surface compared with the Smooth surface produces asperity contact regions that support
a substantial fraction of the external load before global slip begins. The inferred contact state
is in good agreement with the theoretical prediction discussed in Sec.~4.3.

Fig.~\ref{BLOCK.1Logt.2mu.Smooth.replot.eps}(c) shows that the Sb-R surface exhibits the same
general elastic-loading behavior as the Sb-S surface, but reaches the steady-sliding state more rapidly
because its greater roughness promotes faster squeeze-out and earlier formation of asperity
contacts.

\begin{figure}
\includegraphics[width=1.0\columnwidth]{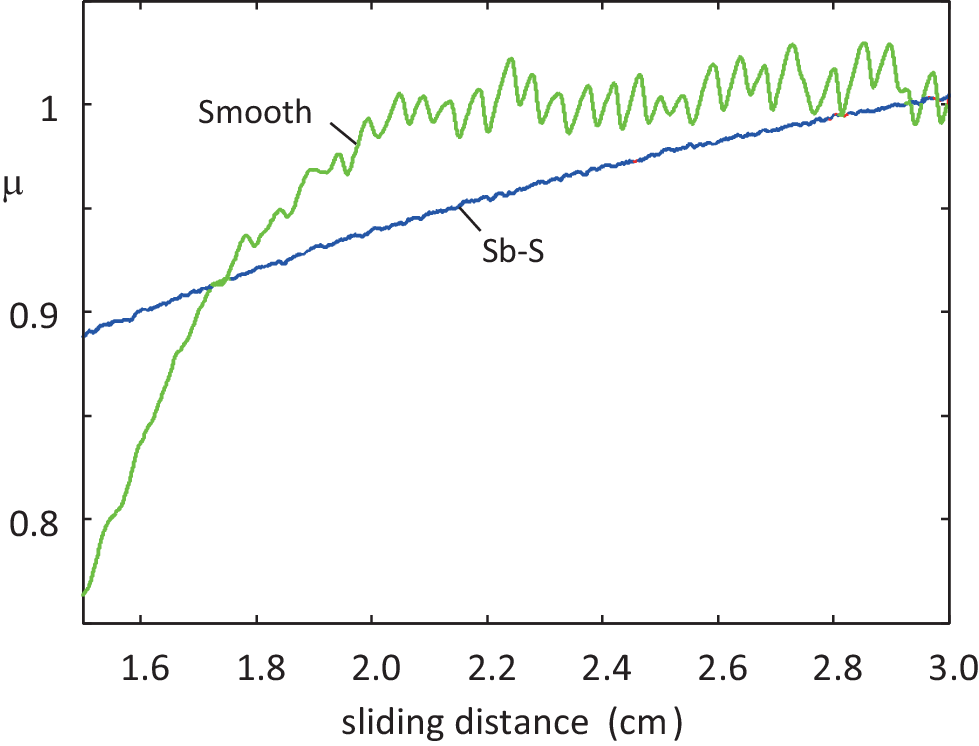}
\caption{\label{1x.2mu.in.glycerol.Smooth.and.Sb-S.Shallamach.eps}
Friction data over the interval $x=1.5$-$3.0 \ {\rm cm}$ from
Fig.~\ref{SCRAPING.1x.2mu.eps}(a) and (b). The friction forces were averaged over
sliding-distance intervals of $0.1 \ {\rm mm}$ to reduce statistical noise. The Smooth surface
exhibits an abrupt onset of frictional instabilities, whereas the sliding motion of the Sb-S
surface remains smooth over the entire sliding-distance range.
}
\end{figure}

Fig.~\ref{1x.2mu.in.glycerol.Smooth.and.Sb-S.Shallamach.eps} shows the friction data over the
interval $x=1.5$-$3.0 \ {\rm cm}$ from Fig.~\ref{SCRAPING.1x.2mu.eps}(a) and (b).
For the Smooth surface, pronounced oscillations begin when the friction reaches a critical value.
These oscillations are consistent with adhesion-induced Schallamach-wave instabilities, as
discussed in Ref.~\cite{XuPersson}. By contrast, the sliding motion of the Sb-S surface remains
smooth, consistent with the suppression of macroscopic adhesion by surface roughness.

In the absence of this transition in sliding mode, the theory predicts that the friction
coefficient of the Smooth surface would reach $\mu\approx1.5$ at
$v=3 \ {\rm \mu m/s}$ and an even larger value, approximately $2.7$, at much lower sliding
speeds; see Ref.~\cite{XuPersson}.

\vskip 0.1cm
\subsection{4.3 Comparison with stationary squeeze-out theory}

We now calculate how surface roughness influences fluid squeeze-out during stationary contact
and show that the theoretical predictions are consistent with the observed dependence of the
friction coefficient at the onset of sliding on the preceding waiting time.

\begin{figure}
\includegraphics[width=0.47\textwidth,angle=0.0]{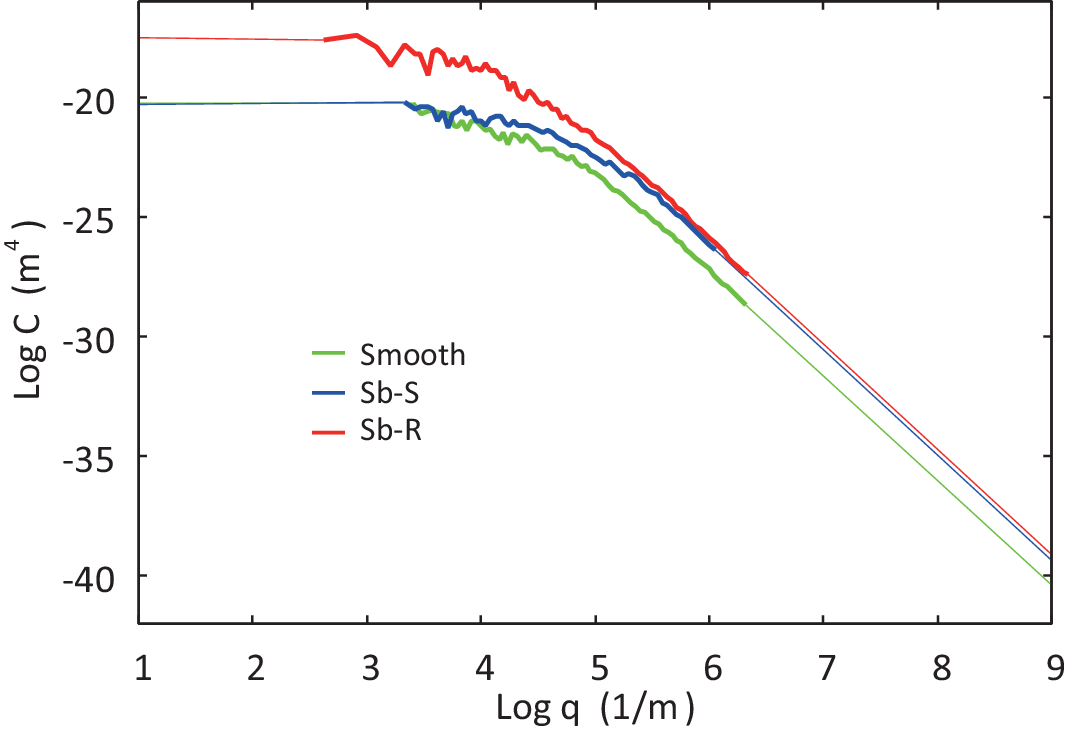}
\caption{\label{cut1.and.not.logq.2logC.all.full.eps}
The thick curves show the measured surface roughness power spectra as functions of wavenumber
on a log-log scale for the Smooth cylinder (green), the Sb-S cylinder (blue), and the Sb-R
cylinder (red). The thin curves indicate the extrapolated regions at small and large wavenumbers.
}
\end{figure}

\begin{figure}
\includegraphics[width=1.0\columnwidth]{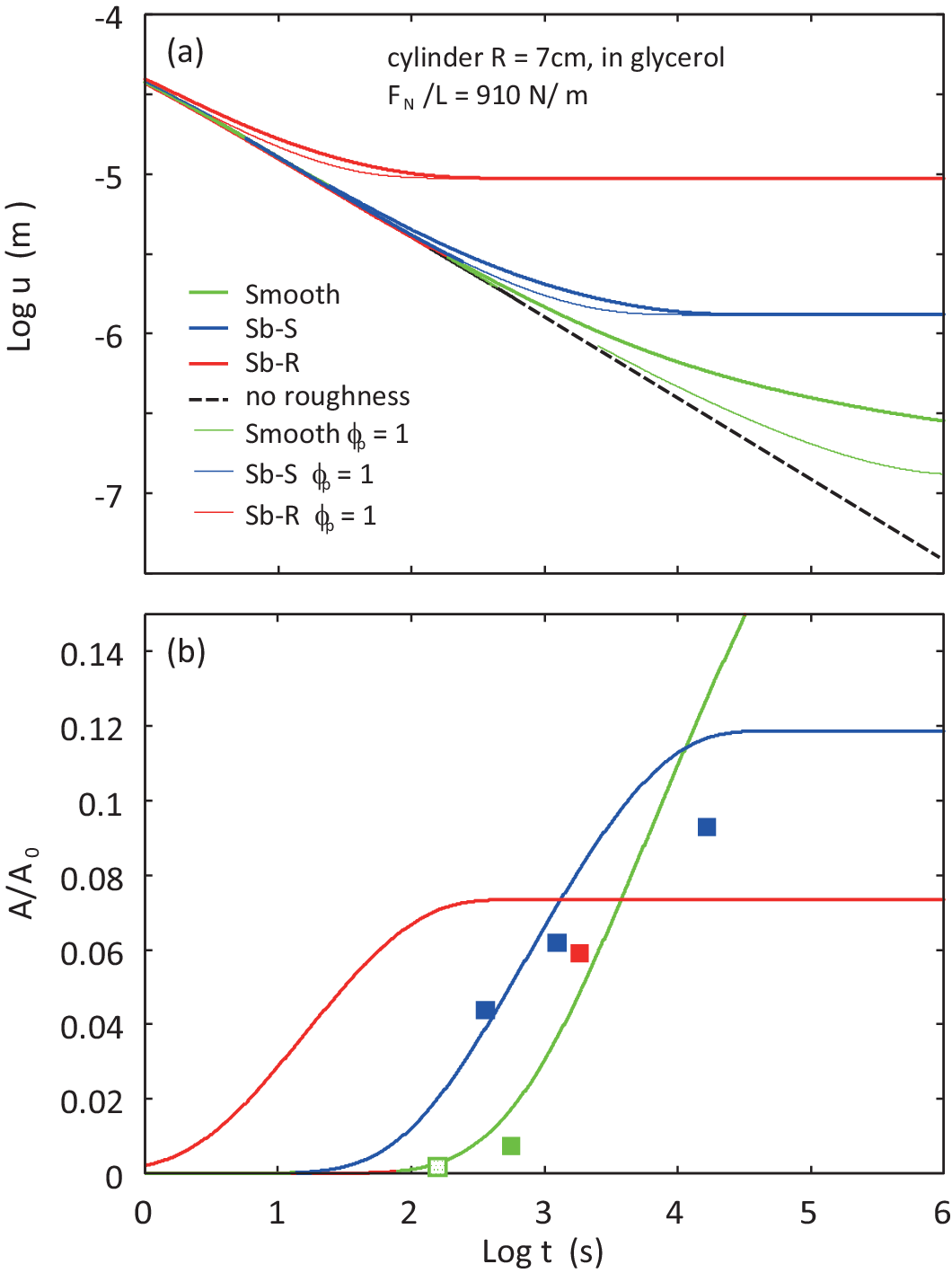}
\caption{\label{1logTime.2logSeparation.squeeze.2plates.eps}
The logarithm of the mean surface separation $\bar u$ in (a) and the relative contact area
$A/A_0$ in (b) as functions of the logarithm of time for the Smooth (green curves),
the Sb-S (blue curves), and the Sb-R (red curves). In (a), the thick curves
were obtained using the full theory, whereas the thin curves were calculated with the pressure
flow factor set to $\phi_{\rm p}=1$. The black dashed curve shows the result for perfectly
smooth surfaces. The calculations were performed using $F_{\rm N}/L=910 \ {\rm N/m}$,
$w=1.27 \ {\rm cm}$, Young's modulus $E=1.5 \ {\rm MPa}$, Poisson ratio $\nu=0.5$,
and fluid viscosity $\eta=1.4 \ {\rm Pa\,s}$.
}
\end{figure}

The thick solid lines in Fig.~\ref{cut1.and.not.logq.2logC.all.full.eps} show the 2D surface
roughness power spectra $C(q)$ calculated from the measured height profiles of the Smooth PMMA
cylinder (green) and the sandblasted Sb-S (blue) and Sb-R (red) cylinders. The measured power
spectra were extrapolated toward both smaller and larger wavenumbers in order to include all
roughness length scales relevant to the contact-mechanics and fluid-flow calculations. These
extrapolated regions are shown by the thin lines in
Fig.~\ref{cut1.and.not.logq.2logC.all.full.eps}.

Including both the measured and extrapolated parts of the power spectra gives the root-mean-square
roughness amplitudes of $h_{\rm rms}=1.37$, $2.80$, and $15.76 \ {\rm \mu m}$ for the Smooth,
Sb-S, and Sb-R cylinders, respectively. The corresponding rms slopes are $0.147$, $0.480$,
and $0.778$.

We next investigate how surface roughness influences fluid squeeze-out during stationary contact
and compare the theoretical predictions with the measured dependence of friction at the onset
of sliding on the waiting time. Fig.~\ref{1logTime.2logSeparation.squeeze.2plates.eps}(a) shows
the calculated mean surface separation for the Smooth, Sb-S, and Sb-R as a function of
the logarithm of time.

The calculations were performed using the plate approximation, which gives nearly the same result
as the full cylinder-contact calculation \cite{Persson1}. In this approximation, the roughness
asperities deform elastically, but the curved cylinder is replaced by a rigid flat plate having
the same width $w$ as the Hertzian contact width of the cylinder-rubber contact. It was shown in
Ref.~\cite{Persson1} that this approximation describes squeeze-out accurately while being
computationally faster and less susceptible to the numerical instabilities that can occur in the
full cylinder-contact calculation.

The thick green, blue and red curves in
Fig.~\ref{1logTime.2logSeparation.squeeze.2plates.eps}(a) were obtained using the full theory,
whereas the thin curves were calculated by setting the pressure flow factor to
$\phi_{\rm p}=1$. The black dashed curve shows the result for a perfectly smooth surface.
For the Sb-R, the fluid is squeezed out over a relatively short time scale of approximately
$100 \ {\rm s}$, whereas for the Sb-S the corresponding time scale is approximately
$10^4 \ {\rm s}$. In the final state, the externally applied load is carried entirely by the
asperity contact regions.

For the sandblasted Sb-S and Sb-R, the full-theory predictions are nearly identical to
those obtained with $\phi_{\rm p}=1$. This is not the case for the Smooth. When
$\phi_{\rm p}=1$, the asymptotic large-time surface separation is reached at approximately
$t=10^6 \ {\rm s}$. In the full theory, however, the final surface separation is not reached
even after $10^{10} \ {\rm s}$, corresponding to approximately $330$ years. This very slow
late-stage squeeze-out results from the strong reduction in effective fluid-flow conductivity
as the non-contact channels approach the percolation threshold.

Fig.~\ref{1logTime.2logSeparation.squeeze.2plates.eps}(b) shows the relative real contact area
as a function of the logarithm of time for the Smooth (green), Sb-S (blue), and Sb-R (red)
surfaces. Consider first the Smooth surface. Experimentally, after a waiting time of
$10 \ {\rm s}$, sliding begins without a detectable elastic-loading stage, and the interface
initially slips with negligible friction. This behavior is consistent with the green curve in
Fig.~\ref{1logTime.2logSeparation.squeeze.2plates.eps}(b), which predicts a negligible real
contact area for $t<100 \ {\rm s}$ and a finite but still very small contact area at
$t\approx150 \ {\rm s}$, as indicated by the open square. At approximately this time, the
contact temporarily enters a sticking or partial-slip state
[see Fig.~\ref{BLOCK.1Logt.2mu.Smooth.replot.eps}(a)]. As the interfacial state continues to
evolve, global sliding resumes at approximately $t=500 \ {\rm s}$.

The filled squares in Fig.~\ref{1logTime.2logSeparation.squeeze.2plates.eps}(b) indicate the
relative contact area for different surfaces and different times, calculated from
$${A\over A_0}=\mu{ p_0\over\sigma_{\rm f}}, \eqno(6)$$
using the measured friction coefficient, and the frictional shear stress $\sigma_{\rm f}$ 
for Smooth, Sb-S and Sb-R surfaces, obtained by fitting the measured Stribeck curve;
see Ref.~\cite{XuPersson}.

The green filled square in Fig.~\ref{1logTime.2logSeparation.squeeze.2plates.eps}(b)
was obtained from (6) for the Smooth surface using
$\sigma_{\rm f}=0.5 \ {\rm MPa}$ from Ref.~\cite{XuPersson} and the friction coefficient
from Fig.~\ref{BLOCK.1Logt.2mu.Smooth.replot.eps} at the onset of the second sliding phase,
at $t\approx500 \ {\rm s}$.

For the Sb-S cylinder, the blue squares show the relative contact areas
calculated at the onsets of slip indicated by the arrows ${\bf a}$, ${\bf b}$, and ${\bf c}$
in Fig.~\ref{1Logt.2mu.NEW3cm.Sb-S.glycerol.2plates.eps}. These values were again obtained
using (6), together with the measured friction coefficients and the frictional shear stress
$\sigma_{\rm f}\approx0.78 \ {\rm MPa}$ obtained from fitting the friction master curve of the
Sb-S cylinder.

The red square shows the corresponding relative contact area for the Sb-R cylinder at the onset of slip
(for $t \approx 1800 \ {\rm s}$). In this case,
we used $\sigma_{\rm f}\approx1.08 \ {\rm MPa}$, as obtained from the fit to the friction master
curve reported in Ref.~\cite{XuPersson}.

Overall, the experimental observations of the onset and evolution of friction are in reasonable agreement with the calculated time dependence of relative contact area for the Smooth, Sb-S, and Sb-R surfaces. This agreement supports the interpretation that the
waiting-time dependence of the friction coefficient at the onset of global sliding results primarily
from the increase in real contact area during fluid squeeze-out [process (a) in Fig. \ref{Scraping.eps} and \ref{BallSlidingPicture.eps}]. 
The results indicate that, at $v=3 \ {\rm \mu m/s}$, the transient dynamic scraping has only a minor influence on the squeeze-out rate.

%%%%%%%%%%%%%%%%%%%%%%%%%%%%%%%%%%%%%%%%%%%%%

\section{5 Summary and conclusion}

We have studied transient fluid removal in cylinder-on-flat contacts formed between PMMA
cylinders and a soft PDMS substrate. In contrast to the block-on-flat geometry, this configuration
does not permit scraping caused by bending of a rubber block. It therefore
allows the influence of sliding-induced elastohydrodynamic fluid removal to be examined more
directly.

For glycerol at the low sliding speed $v=3 \ {\rm \mu m/s}$, the initial increase in friction
after the onset of motion of the machine table is governed mainly by tangential elastic deformation of the contact,
although some local preslip may occur before global sliding. After this elastic-loading stage, 
the friction coefficients obtained after different stationary
waiting times are approximately equal to those reached after sliding for the same total contact
time. The subsequent friction evolution also follows approximately the same dependence on total
contact time, despite the different waiting times and sliding distances. Thus, under these
conditions, the interfacial state is governed primarily by the total time elapsed since application
of the normal load rather than by the preceding sliding distance or contact history.

Stationary squeeze-out calculations based on the measured surface roughness power spectra predict
the time evolution of the mean surface separation and the real contact area. 
The relative contact areas inferred from the measured friction coefficients at the onset of global
sliding are in reasonable agreement with the calculated time dependence for the Smooth, Sb-S, and
Sb-R surfaces. This agreement supports the interpretation that the increase in friction with contact
time originates primarily from the growth of the real contact area during fluid squeeze-out.
Together with the observation that the frictional state depends mainly on the total contact time,
the results indicate that essentially the same squeeze-out process occurs during stationary contact
and low-speed sliding. Thus, at the low sliding speed considered here, sliding-induced
elastohydrodynamic effects have only a minor influence on the fluid-removal process.

Surface roughness and adhesion strongly influence the transient friction response. The rougher
Sb-R surface develops asperity contact more rapidly than the Sb-S and Smooth surfaces. The Smooth
surface exhibits macroscopic adhesion, adhesion-controlled dewetting, and frictional instabilities
consistent with Schallamach waves, and its friction evolution therefore cannot be interpreted
solely in terms of fluid squeeze-out.

%%%%%%%%%%%%%%%%%%%%%%%%%%%%%%%%%%%%%%%%%%%%%

\begin{figure}
\includegraphics[width=1.0\columnwidth]{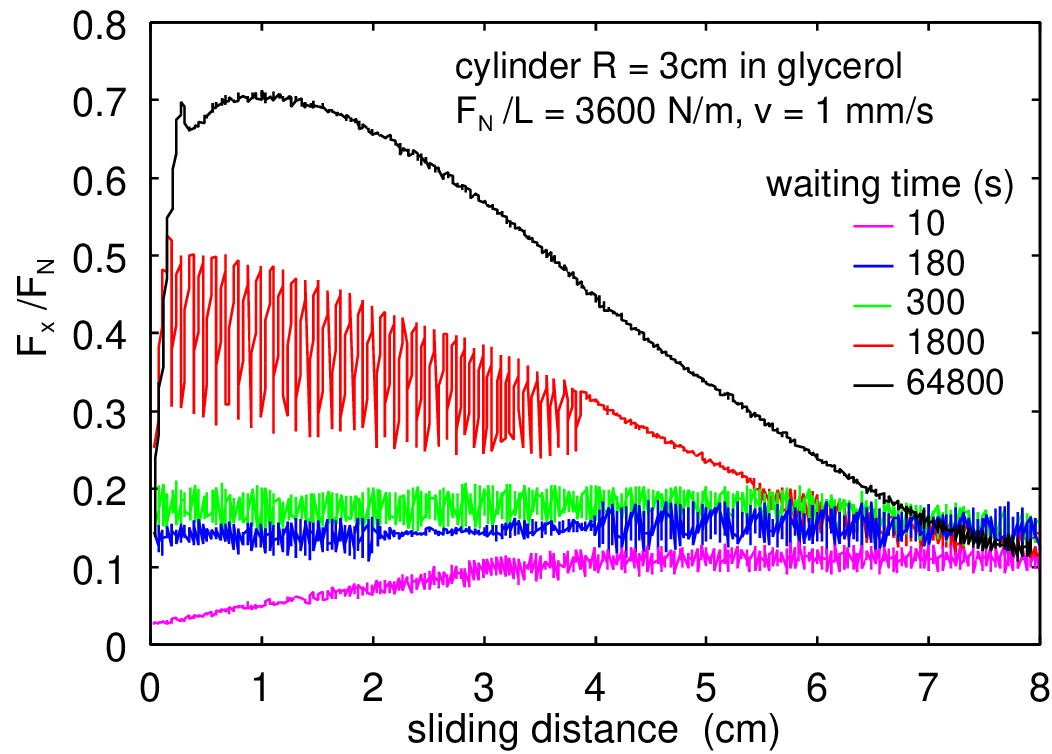}
\caption{\label{BLINK.1x.2mu.6plates.eps}
The ratio $F_x/F_{\rm N}$ as a function of sliding distance for the
$R=3 \ {\rm cm}$ Sb-S cylinder sliding against the silicone rubber surface lubricated with
glycerol. The curves correspond to the different waiting times indicated in the figure.
The normal force per unit length is $F_{\rm N}/L=3600 \ {\rm N/m}$, and the sliding speed is
$v=1 \ {\rm mm/s}$.
}
\end{figure}

\begin{figure}
\includegraphics[width=1.0\columnwidth]{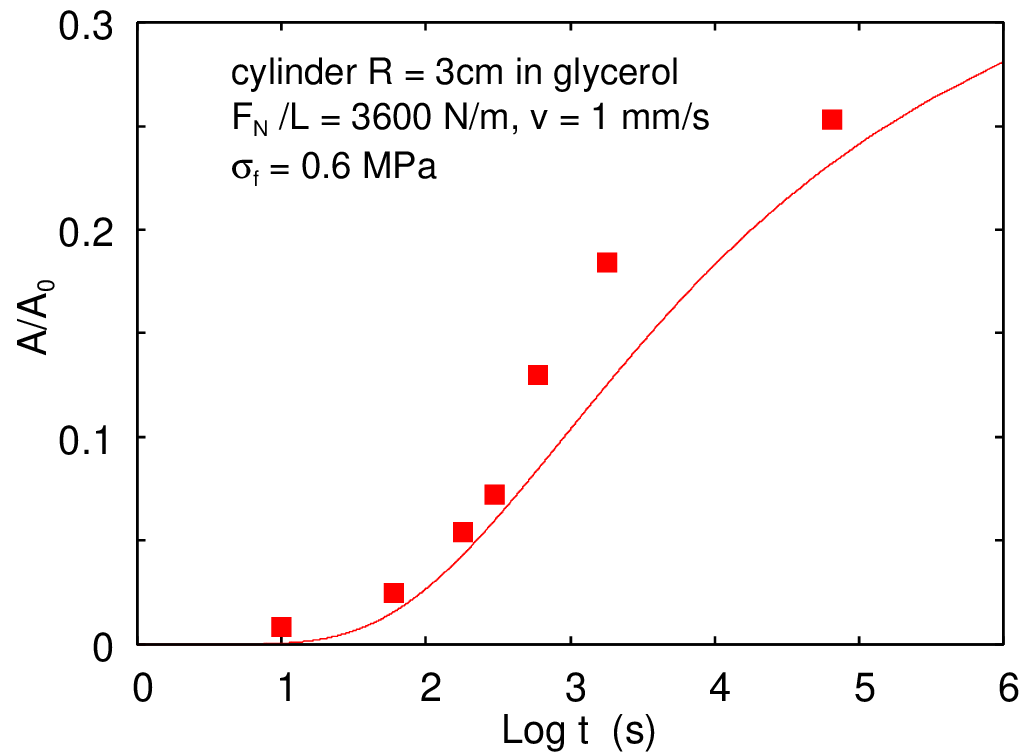}
\caption{\label{BLINK.1Logt.R=3cm.6plates.glycerol.eps}
The relative contact area as a function of the logarithm of waiting time. The solid curve is the
theoretical prediction, and the squares were obtained using (6), with $\mu$ determined from
the friction at the onset of sliding for the data shown in
Fig. \ref{BLINK.1x.2mu.6plates.eps} and two additional cases.
}
\end{figure}

\vskip 0.3cm
\section{Appendix A: Friction of the Sb-S cylinder on silicone rubber in glycerol}

Here, we present additional results for fluid squeeze-out between the Sb-S PMMA surface and the
silicone rubber substrate in glycerol. The applied normal force per unit length was
$F_{\rm N}/L=3600 \ {\rm N/m}$, and the sliding speed was $v=1 \ {\rm mm/s}$.

Fig.~\ref{BLINK.1x.2mu.6plates.eps} shows the ratio $F_x/F_{\rm N}$ as a function of sliding
distance for several waiting times, as indicated in the figure. For the shortest waiting time,
the friction at the onset of sliding is lower than the steady-state friction. By contrast, for the
two longest waiting times, the initial friction exceeds the steady-state value.

Strong stick-slip oscillations occur in some of the experiments. In these cases, we define the
break-loose friction force as the maximum friction force reached during the initial loading and
slip event.

Fig.~\ref{BLINK.1Logt.R=3cm.6plates.glycerol.eps} shows the relative contact area as a function
of the logarithm of waiting time. The solid curve is the theoretical prediction, whereas the squares
were obtained using (6), with $\mu$ determined from the break-loose friction force measured
for the data shown in Fig. \ref{BLINK.1x.2mu.6plates.eps} and for two additional cases not included
in that figure.

In converting the friction coefficients into relative contact areas, we used
$\sigma_{\rm f}=0.6 \ {\rm MPa}$, which is close to the frictional shear stresses obtained
from fitting the Stribeck curves in Ref. \cite{XuPersson} for a larger
cylinder with $R=7 \ {\rm cm}$, but the same surface roughness as the cylinder considered here.

%%%%%%%%%%%%%%%%%%%%%%%%%%%%%%%%%%%%%%%%%%%%%%%%%%%
\vskip 0.3cm
\section{Appendix B: Hydrodynamic re-entrainment in mud}

\begin{figure}
\includegraphics[width=0.47\textwidth,angle=0.0]{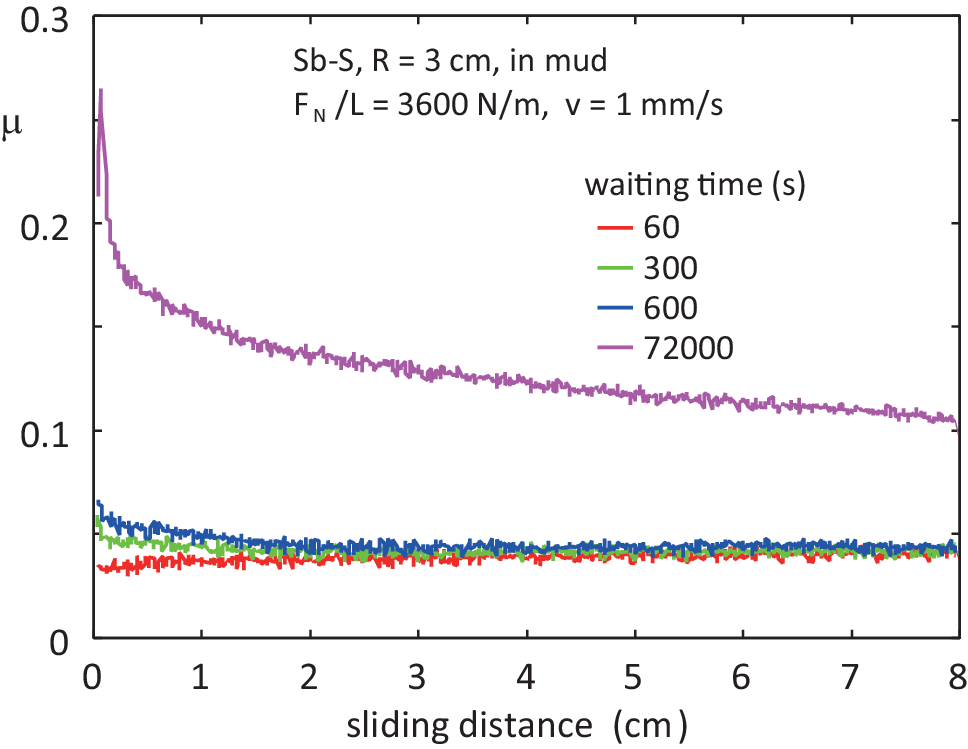}
\caption{\label{1x.2mu.mud.1.5.10.min.and.20hours.eps}
The friction coefficient as a function of sliding distance for the Sb-S PMMA cylinder sliding
in mud. The red, green, blue, and purple curves correspond to waiting times of
$1 \ {\rm min}$, $5 \ {\rm min}$, $10 \ {\rm min}$, and $20 \ {\rm h}$, respectively.
The normal force per unit length is $F_{\rm N}/L=3600 \ {\rm N/m}$, and the sliding speed is
$v=1 \ {\rm mm/s}$.
}
\end{figure}

We next consider the Sb-S PMMA cylinder with radius $R=3 \ {\rm cm}$ sliding in mud. In contrast
to the low-speed glycerol experiments ($v=3 \ {\rm \mu m/s}$), the higher sliding speed $v=1 \ {\rm mm/s}$ used here produces significant
hydrodynamic re-entrainment during sliding.
Fig.~\ref{1x.2mu.mud.1.5.10.min.and.20hours.eps} shows the friction coefficient as a function
of sliding distance at a normal force per unit length of $F_{\rm N}/L=3600 \ {\rm N/m}$. The red, green, blue, and purple curves correspond to waiting times of
$1 \ {\rm min}$, $5 \ {\rm min}$, $10 \ {\rm min}$, and $20 \ {\rm h}$, respectively.

The mud behaves as a highly viscous, non-Newtonian, shear-thinning fluid. At the relatively high
sliding speed used in this experiment, the fluid-film thickness rapidly approaches its dynamic
steady-state value. After a waiting time of $1 \ {\rm min}$, the initial interfacial state is
already close to this steady-sliding state, and the friction coefficient therefore remains nearly
independent of sliding distance.

For longer waiting times, stationary squeeze-out reduces the initial fluid-film thickness below
its steady-sliding value. Hydrodynamic re-entrainment during subsequent sliding then increases the
film thickness, causing the friction coefficient to decrease asymptotically toward its steady-state
value. After the longest waiting time of $20 \ {\rm h}$, the initial film is particularly thin,
and a comparatively long sliding distance is required for the contact to return to the steady-state
sliding condition.

%%%%%%%%%%%%%%%%%%%%%%%%%%%%%%%%%%%%%%%%%%%%%

\begin{figure}
\includegraphics[width=1.0\columnwidth]{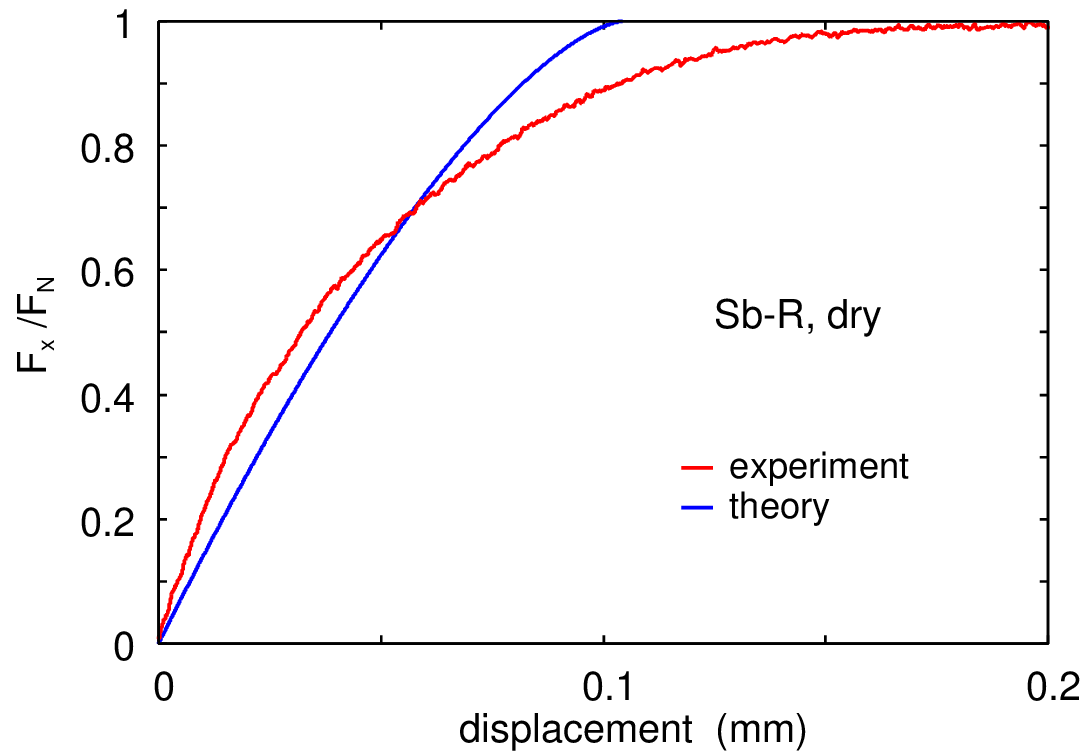}
\caption{\label{1ux.2Fforcemu.eps}
The friction coefficient as a function of the tangential displacement $u_x$ for dry contact
between the Sb-R PMMA cylinder and the PDMS substrate. The normal force per unit length is
$F_{\rm N}/L=310 \ {\rm N/m}$.
}
\end{figure}

\vskip 0.3cm
\section{Appendix C: Contact stiffness}

The relation between the tangential force $F_x$ and the tangential displacement $u_x$ has been
studied for crossed-cylinder contacts \cite{cross}. The infinitely long cylinder-flat geometry
is obtained formally in the limiting case where the radius of one cylinder tends to infinity,
$R\rightarrow\infty$, and the angle between the cylinder axes tends to zero,
$\theta\rightarrow0$.

A cylinder of finite length $L$ cannot, however, be mapped exactly onto the crossed-cylinder
geometry. Nevertheless, we use this mapping to estimate the tangential contact stiffness because
we are not aware of a corresponding analytical theory for a finite-length cylinder-flat contact.

For crossed cylinders, the contact region is elliptical, with a major-axis length $2a$ in the
$x$ direction and a minor-axis length $2b$ in the $y$ direction. We set
$2a=L$ and $2b=(4/\pi)w$, where $w$ is the average width of the ellipse in the $y$ direction,
which is identified with the width of the actual approximately rectangular contact region in the
sliding direction.

Assuming that the local relation
$\tau({\bf x})=\mu p({\bf x})$ applies and that the static and kinetic friction coefficients are
equal, the relation between $F_x$ and $u_x$ for crossed cylinders is
$$
u_x={3\mu F_{\rm N}K\over\pi GL}
\left[
1-\left(1-{F_x\over\mu F_{\rm N}}\right)^{2/3}
\right].
\eqno(A1)
$$
Here,
$$
K=\int_0^{\pi/2}d\phi\,
{1-\nu\sin^2\phi\over
\left(1-e^2\sin^2\phi\right)^{1/2}},
\eqno(A2)
$$
where
$$
e=
\left[
1-\left({b\over a}\right)^2
\right]^{1/2}.
\eqno(A3)
$$

For a sphere-flat contact, $a=b$, and (A1) reduces to the classical Cattaneo-Mindlin
relation \cite{Cat,Mind}. In the limit $F_x/F_{\rm N}\ll1$, (A1) reduces to
$$
u_x\approx{2K\over\pi G}{F_x\over L}.
\eqno(A4)
$$

We compare the prediction of (A1) with measurements for dry contact between the Sb-R PMMA
cylinder and the PDMS substrate at $F_{\rm N}/L=310 \ {\rm N/m}$. This system exhibits no
macroscopic adhesion, although adhesion still increases the asperity contact area.

Using (A2) with
$b/a=(4/\pi)(w/L)=0.094$ gives $K=2.35$. Using this value together with $\mu=1.0$, the blue curve
in Fig. \ref{1ux.2Fforcemu.eps} shows the predicted ratio $F_x/F_{\rm N}$ as a function of the
tangential displacement $u_x$. The red curve shows the corresponding experimental data from
Fig. \ref{1SlidingDistance.2Mu.smooth.dry.lubricated.eps}.

In the limit $F_x/F_{\rm N}\ll1$, the theory predicts a smaller tangential stiffness than is
observed experimentally. In addition, the displacement range required to reach the steady-state
friction coefficient is smaller in the theory than in the experiment. The latter discrepancy may
be related to partial slip before the onset of global sliding.

The nominal contact geometry of the crossed-cylinder contact differs from that of the
cylinder-flat contact. The crossed-cylinder contact is elliptical, whereas the cylinder-flat
contact is approximately rectangular except near the ends of the cylinder, where the stress and
contact geometry are more complex.

A further approximation is the assumption that the static, or break-loose, friction coefficient
$\mu_{\rm s}$ is equal to the kinetic friction coefficient $\mu_{\rm k}$. However, based on the
known influence of $\mu_{\rm s}>\mu_{\rm k}$ on the relation between $u_x$ and $F_x$ for a
sphere-flat contact \cite{Ciava}, this assumption is unlikely to explain the discrepancy between
the theoretical prediction and the experimental results.

\end{document}